\documentclass[prd,nofootinbib,floatfix,amsfonts]{revtex4}

\usepackage{graphicx}
\usepackage{latexsym}

\begin{document}
\title{Constraints on CPT violation from WMAP three year polarization data: a wavelet analysis}

\author{Paolo Cabella$^a$}
\email{cabella@astro.ox.ac.uk}

\author{Paolo Natoli$^b$}
\email{paolo.natoli@roma2.infn.it}

\author{Joseph Silk$^a$}
\email{silk@astro.ox.ac.uk}

\affiliation{ $^a$ University of Oxford, Astrophysics, Keble Road, Oxford, OX1 3RH, U.K.}

\affiliation{ $^b$ Dipartimento di Fisica e sezione INFN, Universit\`a di Roma ``Tor Vergata'', Via della Ricerca Scientifica, I-00133 Roma, Italy}

%\date{\today.}

\begin{abstract}
  We perform a wavelet analysis of the temperature and polarization
  maps of the Cosmic Microwave Background (CMB) delivered by the WMAP
  experiment in search for a parity violating signal. Such a signal
  could be seeded by new physics beyond the standard model, for which
  the Lorentz and $\mathcal{CPT}$ symmetries may not hold. Under these
  circumstances, the linear polarization direction of a CMB photon may
  get rotated during its cosmological journey, a phenomenon also
  called cosmological birefringence.  Recently, Feng et al.\ have
  analyzed a subset the WMAP and BOOMERanG 2003 angular power spectra
  of the CMB, deriving a constraint that mildly favors a non zero
  rotation.  By using wavelet transforms we set a tighter limit on
  the CMB photon rotation angle $\Delta\alpha= -2.5\pm 3.0 $
  ($\Delta\alpha= -2.5\pm 6.0 $) at the one (two) $\sigma$ level,
  consistent with a null detection.
\end{abstract}

\maketitle
\section{Introduction}
The CMB is one of the primary experimental windows to the early
universe. Recent observations have reached remarkable precision. When
combined with other complementary cosmological datasets, the WMAP
three year (hereafter, WMAP3) observations \cite{lambda} 
convincingly support the so-called standard model of structure
formation~\cite{spergel}. However the CMB may provide further
information. In principle, one may use the background photons to
constrain new physics beyond ``standard'' models. A positive answer
might be provided by the study of CMB polarization
(CMBP), whose observations currently mark the experimental frontier of
the field.  Pioneering observations, including DASI
\cite{kovac}, CBI \cite{readhead}, BOOMERanG 2003
\cite{piacentini-montroy} (hereafter B03), MAXIPOL \cite{johnson}, and
WMAP itself \cite{page} have yielded detections of the CMBP over a
wide range of angular scales.  Within the next decade, ground or
space-based experiments may detect via the CMBP a signal from
primordial gravitational waves \cite{kks-seljak}, thus constraining
the energy scale of the inflation and probing particle physics well
beyond the capability of any conceivable terrestrial accelerator.  The
CMBP can also provide information on symmetry-violating physics beyond
the Lee-Yang parity ($\mathcal{P}$) breaking that is central to the
standard model, yet not observable through CMB anisotropies due to
their charge blind character. In general, the breakdown of spacetime
symmetries is a potential tracer of new physics \cite{lehnert}.
Several models exist that predict non-standard $\mathcal{P}$ and
$\mathcal{CP}$ violations ('$\mathcal{C}$' standing for charge
conjugation), as well as $\mathcal{CPT}$ violations ('$\mathcal{T}$'
being time reversal) and the related (through the anti-$\mathcal{CPT}$ theorem
\cite{greenberg}) breakdown of Lorentz invariance. A number of
tests have been suggested and (in many cases) performed, either in
terrestrial and orbital laboratories \cite{bluhm-mewes} or through
cosmological observations \cite{amelino,carroll90}.  These violations
might also have a measurable imprint on the observed CMBP pattern,
whose statistical properties are constrained by the assumption of
symmetry conservation.
  
For a sky direction $\hat{n}$, a polarized map of the CMB is usually
given in terms of total intensity (or temperature) $T(\hat{n})$ and
linear polarization Stokes parameters $Q(\hat{n})$ and
$U(\hat{n})$. The $T$ field can be decomposed into scalar (S)
spherical harmonics $Y_{lm}(\hat{n})$, obtaining the coefficients
$a^T_{lm}$. $Q$ and $U$ are components of a symmetric, trace-free rank
2 tensor, and are expanded in tensor spherical harmonics
$Y^G_{lm}(\hat{n})$ and $Y^C_{lm}(\hat{n})$ with coefficients
$a^G_{lm}$ and $a^C_{lm}$, respectively. These correspond to scalar
(gradient-like) '$G$' and pseudo-scalar (curl-like) '$C$' modes
\cite{kks-seljak}.  Under hypothesis of Gaussianity and
isotropy, the statistical properties the CMB are described by two
point correlation functions on the sphere, whose Legendre transforms
define six angular power spectra: $C^{ZZ'}_l=\langle
a^{Z}_{lm}(a^{Z'}_{lm})^*\rangle$ with $Z,Z'=\{T,G,C\}$.  If the
physics controlling CMB fluctuations is parity conserving the cross
spectra $C^{TC}_l$ and $C^{GC}_l$ must vanish due to the different
handedness of the C and (S,G) harmonics. Therefore, if the standard
cosmological model holds ,we should expect no relevant information
from $TC$ and $GC$. On the other hand, detection of non-zero
primordial $TC$ and/or $GC$ may probe fundamental physics in the early
universe, such as the presence of a primordial homogeneous
\cite{scafer} or helical \cite{pogosian} magnetic field which would induce
Faraday rotation and non-zero $TC$ correlations. Parity-asymmetric
gravity dynamics during inflation may generate a discrepancy among
left and right-handed gravitational waves, so that $TC$ and $GC$ are
non-zero \cite{lue}. Particle physics models with non-standard
parity-violating interactions also predict non-vanishing $TC$ and $CG$
signals \cite{maity}.

In this paper we focus on a class of models that exhibit parity
violations in  the photon sector. A Chern-Simons term is
introduced in the effective Lagrangian \cite{carroll90}:
\[
  \Delta{\mathcal L}= -\frac{1}{4}\,  p_{\mu} 
  \epsilon^{\mu\nu\rho\sigma}F_{\rho\sigma} A_{\nu}\;,   
\]
where ${F}^{\mu\nu}$ is the Maxwell tensor and $A^\mu$ the 4-potential.
The 4-vector $p_{\mu}$ may be interpreted as the derivative of the
quintessence field or the gradient of a function of the Ricci scalar
\cite{Davoudiasl}. In either case a $\mathcal{P}$ violation always arises provided
that $p_0$ is non-zero, while $\mathcal{C}$ and $\mathcal{T}$ remain
intact. Hence, $\mathcal{CP}$ and $\mathcal{CPT}$ symmetries are also
violated, as well as Lorentz invariance, since $p^\mu$ picks up a
preferred direction in space-time. The net effect on a propagating
photon is to rotate its polarization direction by an angle
$\Delta\alpha$, hence the name ``cosmological birefringence''.
Historically, the effect has being constrained by measuring polarized
light from high redshift radio galaxies and quasars
\cite{carroll90,carroll97-nodland}.  Obviously, the CMB photons would
also be affected and, due to their longer journey, may get a larger
rotation. A consequence for the CMB pattern is the mixing of $G$ and $C$
modes: the $TG$ and $GC$ correlations still vanish at last scattering
surface, but the observable CMB spectra are distorted as
\cite{lue,fllz}:
\begin{eqnarray}
\label{eq:TC}
C_l'^{TC}&=& C_l^{TG}\sin 2 \Delta\alpha\\
\label{eq:GC}
C_l'^{GC}&=&\frac{1}{2}(C_l^{GG}- C_l^{CC})\sin 4 \Delta \alpha \\
\label{eq:TG}
C_l'^{TG} &=& C_l^{TG}\cos 2 \Delta \alpha \\
\label{eq:GG}
C_l'^{GG} &=& C_l^{GG}\cos^2 2 \Delta \alpha + C_l^{CC}\sin^2 2 \Delta \alpha\\
\label{eq:CC}
C_l'^{CC} &=& C_l^{CC}\cos^2 2 \Delta \alpha + C_l^{GG}\sin^2 2\Delta \alpha.
\end{eqnarray}
where the primed quantities are rotated. In \cite{feng}, the $TT$ and
$TG$ power spectra measured by WMAP3 together with all six spectra
measured by B03 have been used to perform a global fit, yielding a
mild detection for a non zero rotation (but see also \cite{kost} for a
similar analysis restricted to the B03 power spectra and \cite{liu}
where constraints on the coupling between the quintessence and the
psudoscalar of electromagnetism are derived, based again on B03 data). Using the same data set, a similar result has been found
in \cite{li}, and used to constrain a specific baryo/leptogenesis
model, while an interaction between the neutrino asymmetry and a term
Chern-Simons term has been proposed in \cite{geng} as a possible
explanation for the result found \cite{feng}. Forecasted
  constraitns on $\Delta \alpha$ for high sensitivity experiments such
  as Planck or CMBpol can be found in \cite{Xia2}.

Here we constrain $\Delta\alpha$ with a wavelet analysis. A rotation
of the photon polarization direction leaves an imprint on each
resolution element (or pixel) of the Q and U maps, and a map-based
estimator appears appropriate. Wavelets are a natural choice because
they allow for multi-scale pixel analysis. We compute the wavelet
cross-correlation coefficients for $TC$ and $GC$ to build a goodness
of fit estimator that we apply to the WMAP3 $\{ T,Q,U\}$ maps.  Our
analysis is complementary to that of \cite{feng}, where the
information for $TC$ and $GC$ comes from B03.  The two analyses differ
in the method and (substantially) in the data set (the only overlap
being the WMAP3 temperature map).  In the following, we derive more
stringent limits on $\Delta\alpha$ by adapting the wavelet formalism
to tackle polarization, a point that has not been addressed to date in
the CMB literature. 

\noindent  The plan of the paper is as follows: in section
  \ref{sect:method} we describe our wavelet based method to constrain
  $\Delta\alpha$: section \ref{sect:sim} we define a suitable
  estimator and apply it to WMAP3 data, making use of
  numerical simulations. Finally in section \ref{sect:concl} we draw
  our conclusions.

\section{A wavelet statistic for temperature and polarization}
\label{sect:method}
Given a position $\vec X$, wavelets are filter functions
$\Psi(\vec{X};b,R)$ that also depend on a characteristic scale $R$ and
translation $b$. They provide scale-varying transforms that remain
localized in pixel space.  Moreover, they consist of an infinite set
of basis functions, thus providing some freedom of choice in matching
their functional form to the target signal. Several authors have
exploited this flexibility as a powerful tool in CMB data analysis.
Wavelets have been used for denoising \cite{sanz}, point source
extraction \cite{cayon}, foreground removal \cite{frode} and for detecting 
the integrated Sachs Wolfe effect \cite{needlets}. Since the
wavelet transform preserves linearity, its coefficients can be used to
constrain the statistics of the field at different scales. In
particular, the spherical Mexican hat wavelet (SMHW) has been used to
flag statistical anomalies in the WMAP data \cite{vielva} and to
constrain primordial non-Gaussianity \cite{mw} (other types of
wavelets have been shown to be sensitive to yet different anomalies,
e.g.\cite{mcewen}). SMHW are generated from ordinary MH wavelets
through a stereographic projection on the tangent plane
\cite{martinez} that is known to preserve their basic properties
\cite{antoine}.  The SMHW is defined as:
\[
\Psi(y,R)=\frac{1}{\sqrt{2\pi}N(R)}\left[1+(\frac{y}{2})^2\right]^2\left[2-(\frac{y}{R})^2e^{-y^2/2R^2}\right]
\]
where $y=2\tan\theta/2$ ($\theta$ is the polar angle), $R$ is the
scale of convolution, and $N(R)$ a normalization factor.  For a T map,
the wavelets coefficients are:
\[
W^T(R,\hat{n})=\int d\Omega'T(\hat{n}+\hat{n}')\Psi(\theta',R)\; .
\]
This convolution can be performed in harmonic space:
\begin{equation}
\label{eq:wavecoeff}
W(R,\hat{n})=\sum_{lm}(\frac{4\pi}{2l+1})^{1/2}a^T_{lm}\Psi_l(R)Y_{lm}(\hat{n})
\end{equation} 
where $\Psi_l(R)$ are the Legendre expansion coefficients of the SMHW.
Handling polarization requires more care, since $Q$ and $U$ are not
rotationally invariant, being components of the rank 2 tensor $P_{ab}$
\cite{kks-seljak}. By taking the covariant derivatives of $P_{ab}$, one
can build two quantities that are rotational invariant and hence
decomposed by S harmonics. This leads, again, to the $G$ and $C$
coefficients:
\begin{eqnarray*}
  a_{lm}^G &=&N_l\int d\Omega  P_{ab}^{\::ab}(\hat{n})Y^{*}_{lm}(\hat{n})\\
  a_{lm}^C &=&N_l\int d\Omega P_{ab}^{\::ac}(\hat{n})\epsilon_{c}^{\:b}(\hat{n})Y^{*}_{lm}(\hat{
n}).
\end{eqnarray*}
Here ':' stands for the covariant derivatives on the sphere,
$\epsilon$ is the Levi-Civita trace-free antisymmetric tensor and
$N_l$ a normalization factor \cite{kks-seljak}. We similarly define
SMHW coefficients as:
\begin{eqnarray*}
%\label{eq:coeff}
  W^G(\hat{n},R)&=&\int d\Omega'P_{ab}^{\::ab}(\hat{n}+\hat{n}')\Psi(\theta',R)\\
  W^C(\hat{n},R)&=&\int d\Omega'P_{ab}^{\::ac}(\hat{n}+\hat{n}')\epsilon_{c}^{\:b}(\hat{n}+\hat{n}')\Psi(\theta',R)\; . 
\end{eqnarray*}
Note that we never explicitly compute derivatives, since the integrals
can be performed in harmonic space (c.f.\ \ref{eq:wavecoeff}),
provided we divide out the factor $N_l$. Finally, we consider the
pixel-pixel cross correlation of the SMHW coefficients as our main
statistic:
\begin{equation}\label{eq:intcross}
X^{Z C}(R) = \frac{1}{V}\int W^{Z}(\hat{n},R)W^C(\hat{n},R)d\hat{n}
\end{equation}
where $Z = \{T,G\}$ and $V$ is a volume normalization that can be
taken to be  proportional to the total number of pixels $N_p$.  The
quantities in eq.~\ref{eq:intcross} possess the same P symmetry of the
usual harmonic cross spectra: they can be non-zero only if parity
conservation is violated.
\section{Numerical simulations and results}
\label{sect:sim}
To constrain $\Delta\alpha$, the following scheme was employed. We
modified the Healpix package \cite{healpix} to generate a set of Monte
Carlo (MC) simulations for $\{T,Q,U\}$ maps containing a CMB signal
whose polarization pattern is rotated according to eqs.\
\ref{eq:TC}-\ref{eq:CC}; we use the WMAP3 best fit model as the
(unrotated) fiducial angular power spectrum. The signal maps were
smoothed according to the WMAP3 optical transfer function. We also
simulated noise maps consistent with the WMAP3 instrumental
properties. We add to each signal map a noise realization
consistent with the WMAP3 instrumental properties. Simulating noise in
TQU maps is more complicated than for T only, because the noise values
of different Stokes parameters within a given pixel are usually
correlated. For WMAP3, $T$ is very weakly correlated with $Q$ and $U$,
so this coupling can be safely neglected~\cite{jarosik}. On the
contrary, in order to obtain accurate results one has to
take into account the correlations between $Q$ and $U$. The WMAP team has
released $2\times 2$ effective hits arrays (hereafter,
$N_{\mathrm{obs}}$ where the off diagonal elements represent the
$\langle QU\rangle$ inter pixel correlation (off pixel correlations
are very weak and can be neglected); these matrices are given for each
differential assembly (DA) and for each observation year. To simulate
noise maps, the following scheme is employed: for a given DA and for
each pixel, we add the $\mathbf{N}_{\mathrm{obs}}$ arrays for
different years. The three year, noise maps for each DA $i$ are then
simulated as:
\begin{equation}
\left(
\begin{array}{c}
Q_{i,p}\\
U_{i,p}
\end{array}\right)= \mathbf{N}_\mathrm{obs}^{-1/2}(i,p) \sigma_{QU}(i)
\end{equation}
where $p$ identifies a given pixel, $\sigma_{QU}$ is the nominal DA
polarization sensitivity, as provided by the WMAP team~\cite{jarosik}
and $\mathbf{N}_\mathrm{obs}^{1/2}$ is the Choleski factor of
$\mathbf{N}_\mathrm{obs}$. The resulting noise plus signal maps for
each DA are then weighted averaged to form a combined map comprising
all DA's in the Q,V and W bands:
\begin{equation}
  \left(
\begin{array}{c}
Q_p\\
U_p\end{array}\right) = \left[ \sum_i^{N_{DA}} \mathbf{C}_{ip}^{-1}
\right]^{-1}\sum_i^{N_{DA}}\mathbf{C}_{ip}^{-1}\left(
\begin{array}{c}
Q_{ip}\\
U_{ip}\end{array}\right)
\end{equation}
The combined map for $T$ is computed by using the standard, scalar version
of the procedure above:
\begin{equation}
T_p=\sum_i^{N_{DA}} \frac{T_{ip}}{\sigma^2_i}\left[\sum_i\frac{1}{\sigma^2_i} \right].
\end{equation} 
To minimize residual foreground
contamination we chose to use a rather conservative mask, the
intersection of the Kp0 and P02 sky cuts
\cite{lambda}. The masked maps are downgraded in resolution to
$13.6'$. We then compute the wavelet coefficients $W^Z(p,R)$ (where
$Z=\{T,G,C\}$) over the discretized sphere. We consider 17 wavelet
scales from $14'$ to $100'$.\footnote{The exact set of scales
considered is
$R=[14,16,18,20,22,25,30,35,40,45,50,55,60,65,70,80,90,100]$ arcmin.}
To avoid boundary effects, we widen the sky map up to a fraction
$2.5\,R$ \cite{vielva,mw}. Finally, we consider the goodness of fit
statistics $\chi^2(\Delta\alpha)= Y^T
\mathbf{C}_{\Delta\alpha}^{-1}Y$, where
$Y=X^{\mathrm{WMAP}}-\bar{X}(\Delta\alpha)$. $X^{\mathrm{WMAP}}$ is
computed over the Q+V+W foreground-cleaned, optimally-weighted data
map, while the mean (barred) quantities are derived from $\sim 2000$
MC simulations. The covariance matrix $\mathbf{C}_{\Delta\alpha}$ is
estimated over a fresh set of simulations ($\sim 4000$).

\begin{figure}%[tbp]
\includegraphics[width=6.cm,angle=0]{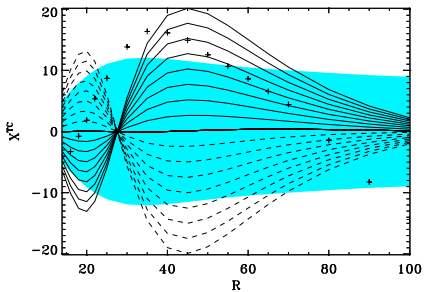}
\caption{MC means for $X^{TC}$ (see text), for
  $-8^\circ\leq\Delta\alpha\leq 8^\circ$, step of $1^\circ$ (dashed
  lines refer to positive $\Delta\alpha$, the middle line is for
  $\Delta\alpha=0$, i.e.\ $\mathcal{P}$ is conserved). The scale $R$ is given in
  arcminutes. The shaded region shows the $1\sigma$ range for
  $\Delta\alpha=0$. Experimental points (WMAP3) are shown as crosses.}
\label{fig:averagesTC} 
\end{figure}

\begin{figure}%[tbp]
\includegraphics[width=6.cm,angle=0]{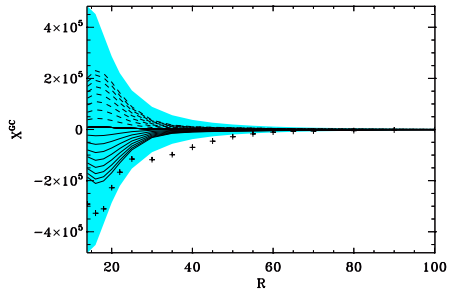}
\caption{Same as fig.~\ref{fig:averagesTC} but for $X^{GC}$ and  
$-16^\circ\leq\Delta\alpha\leq 16^\circ$, step of $2^\circ$. Note the
lower signal to noise.}
\label{fig:averagesGC} 
\end{figure}

In fig.~\ref{fig:averagesTC} we show, as a function of $R$, the MC
mean values for $X^{TC}$, for $-8^\circ\leq\Delta\alpha\leq 8^\circ$
with a step of $1^\circ$.  The crosses are experimental points from
WMAP3 and the shaded region is the $1\sigma$ range, centered about the
$\Delta\alpha=0$ case. In fig.~\ref{fig:averagesGC} we show the same
for $X^{GC}$, with the means in the range
$-16^\circ\leq\Delta\alpha\leq 16^\circ$ with a step of $2^\circ$. As
expected for WMAP3, $TC$ has a significantly larger signal to noise than
GC. The means are computed over noisy simulations, but closely
reproduce the ensemble predictions that can be derived from
eq.~\ref{eq:intcross}.
To show that our estimator is unbiased, we simulated a further MC set
with given $\Delta\alpha$ and checked that the means of the $\chi^2$
estimates reproduce the input values with high accuracy. Throughout
our analysis, we keep the dependence of $\Delta\alpha$ in the
estimator's covariance matrix (but find no significant change in our
results if we drop this dependence: for WMAP3, our estimator's
covariance is completely dominated by noise).

\begin{figure}%[tbp]
 \includegraphics[width=6.cm,angle=0]{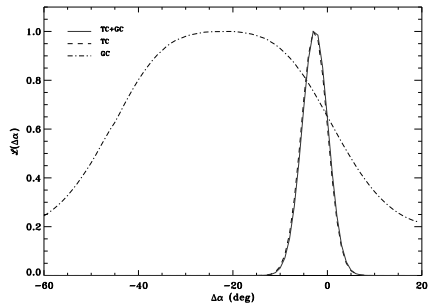}
\caption{Likelihood functions of the cosmological birefringence angle
  $\Delta\alpha$ for CG (dotted) and TC (dashed), computed from the
  wavelet estimator on WMAP3 data. The solid (blue) line is the
  global, covariance weighted, likelihood}
\label{fig:like}
\end{figure}

In fig.~\ref{fig:like} we show the likelihoods of WMAP3 data for $TC$
and $GC$. $GC$ contributes very weakly to the joint likelihood
$\mathcal{L}\propto\exp (-\chi^2/2)$. We estimate $\Delta\alpha=
-2.5\pm 3.0$ and $\Delta\alpha= -2.5\pm 6.0$ at $1\sigma$ and $2\sigma$
confidence limit respectively. Thus, we find no evidence of parity
violation from the WMAP3 data. These limits are slightly tighter than
those given in \cite{feng}, where a marginal detection for a non-zero
$\Delta \alpha$ is claimed and seems to be driven by the GC B03 data.
To show that our conclusions do not depend on the particular fiducial
power spectrum chosen (provided it is reasonable), we have allowed the
latter to vary between the $\pm 1\sigma$ experimental limits set by
WMAP3, finding fully consistent results (no detection, very similar
limits). This procedure extends to polarization the test suggested by
\cite{vielva} for temperature data.

 As a futher consistency check, we compared our estimator with a
similar one, built using the angular power spectrum rather than
wavelets. In the case of pure signal, under the assumption of
Gaussianity and statistical isotropy of the observed field, we expect
the two approaches to provide similar constrains. To show this is
indeed the case, we have repeated the procedure of section
\ref{sect:sim} performing a Monte Carlo simulation over 1000
realizations in the ideal case of pure signal. The $\chi^2$ using the
cross spectrum $C_{l}^{TC}$ can be calculated analytically by:
\begin{equation}
\chi^2(\Delta\alpha) = \sum (C_l^d-C_{l}'^{TC})^2/\sigma_l^2
\end{equation}
where as usual the prime identified rotated spectra, and the cosmic
variance is given by \cite{kks-seljak} :
\[
\sigma_{l}^{2}=[(C_{l}'^{TC})^2+C_{l}^{TT}C_{l}'^{CC}]/2l+1
\]
\begin{figure}%[tbp]
 \includegraphics[width=6.cm,angle=0]{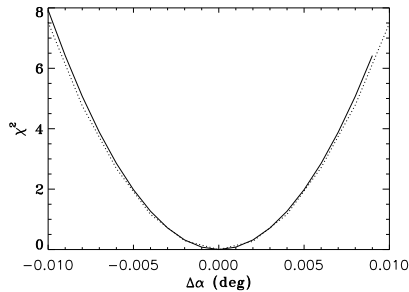}
\caption{$\chi^2$ of the cosmological birefringence angle $\Delta\alpha$ in  case of pure signal using the wavelets estimator of the component TC (dotted) and the angular power spectrum $C_{l}^{TC}$ (solid)}
\label{fig:like_comp}
\end{figure}
and the maximum multipole in the sum is $l_{max}\simeq 500$  roughly consistent with the maximum resolution emploied in the wavelet analysis.
In figure \ref{fig:like_comp} we show $\chi^2$ as a function of
$\Delta\alpha$ against the null hypothesis for wavelets and the cross
spectrum $C_{l}^{TC}$. The two methods give very similar results, as
expected in this ideal case.

\section{Conclusions}

\label{sect:concl}

In summary, we have presented the first application of wavelets to
polarized CMB maps, and used it to constrain the rotation angle of CMB
photons in search of a signature due to cosmological birefringence, an
effect connected to fundamental symmetry-breaking physics.  We find no
evidence of such a rotation and present the best upper limits to date
on CMB data. This result should be compared with \cite{feng}, where a
marginal detection for a non-zero $\Delta\alpha$ is claimed. The
latter result is mostly based on the B03 data and only makes use of a
subset of the WMAP3 dataset, \textit{not} including the $TC$
correlations from which our results are essentially derived. While
WMAP3 has lower signal to noise \textit{per pixel} than B03, the
analysis presented here uses data from $\sim 60\%$ of the whole sky,
while the limited useful sky coverage of B03 ($\lesssim 1\%$) severely
limits the statistical power of $TC$, so the detection in \cite{feng}
appears to be driven from the much harder to measure (and prone to
systematic effects) $GC$ correlations~\footnote{Recently, Xia et
  al. \cite{Xia} have extended the analysis in \cite{feng} including
  in their analysis the previously left aside WMAP3 full power
  spectrum dataset. They find $\Delta\alpha=-6.2\pm3.8 \mathrm{deg}
  (1\sigma)$, thus confirming a mild detection of a non-zero rotation
  angle. It would be very interesting to understand to what extent
  this is driven by the B03 dataset. An extension of our analysis to
  the B03 TQU maps (not yet public at the time of writing) is under
  study.}. Given the quantity and quality of the CMB data anticipated
over the next few years, our approach demonstrates that substantially
stronger limits on parity violation should be feasible.
\section{Acknowledgments}
We thank V.\ Antonuccio, O.\ Dor\'e, M.\ Lattanzi and A.\ Palazzo for
discussions. This research used resources of the NERSC, which is
supported by the Office of Science of the U.S.\ DoE under Contract
No.\ DE-AC02-05CH11231, and of the CASPUR supercomputing centre (Rome,
Italy). Some of the results presented here have been derived using the
Healpix package \cite{healpix}.

\end{document}